\documentclass[twocolumn,showpacs,preprintnumbers,prb,epsfig,superscriptaddress]{revtex4}

\usepackage{graphicx}% Include figure files
\usepackage{dcolumn}% Align table columns on decimal point
\usepackage{bm}% bold math

\usepackage{textcomp} % To write the celsius symbol.
\begin{document}

\preprint{K. W. Kim et al., PRB 81, 214508 (2010)}

\title{Evidence for multiple superconducting gaps in optimally doped
BaFe$_{1.87}$Co$_{0.13}$As$_{2}$ from infrared spectroscopy}

\author{K. W. Kim}
%\email{kyungwan.kim@unifr.ch}

\affiliation{Department of Physics and Fribourg Center for Nanomaterials, University of Fribourg, Chemin
du Mus$\acute{e}$e 3, CH-1700 Fribourg, Switzerland}

\author{M. R\"{o}ssle}
\affiliation{Department of Physics and Fribourg Center for Nanomaterials, University of Fribourg, Chemin
du Mus$\acute{e}$e 3, CH-1700 Fribourg, Switzerland}

\author{A. Dubroka}
\affiliation{Department of Physics and Fribourg Center for Nanomaterials, University of Fribourg, Chemin
du Mus$\acute{e}$e 3, CH-1700 Fribourg, Switzerland}
\author{V. K. Malik}
\affiliation{Department of Physics and Fribourg Center for Nanomaterials, University of Fribourg, Chemin
du Mus$\acute{e}$e 3, CH-1700 Fribourg, Switzerland}

\author{T. Wolf}
\affiliation{%Forschungszentrum Karlsruhe,
Institut f\"{u}r Festk\"{o}rperphysik, 76021 Karlsruhe, Karlsruher
Institut f\"{u}r Technologie, Germany}

\author{C. Bernhard}
\email{christian.bernhard@unifr.ch}

\affiliation{Department of Physics and Fribourg Center for Nanomaterials, University of Fribourg, Chemin
du Mus$\acute{e}$e 3, CH-1700 Fribourg, Switzerland}

%\date{\today}

\begin{abstract}
We performed combined infrared reflection and ellipsometry
measurements of the in-plane optical response of single crystals
of the pnictide high temperature superconductor
BaFe$_{1.87}$Co$_{0.13}$As$_{2}$ with $T_{c}$ = 24.5 K. Our
experimental data provide evidence for multiple superconducting
energy gaps and can be well described in terms of three isotropic
gaps with 2$\Delta/k_{B}T_{c}$ of 3.1, 4.7, and 9.2. The obtained
low-temperature value of the in-plane magnetic penetration depth
is 270 nm.
\end{abstract}


\maketitle

The origin of the recently discovered high temperature
superconductivity (HTSC) in the pnictides is a subject of great
interest.\cite{Kamihara08} In the pnictides, it has been predicted
that up to five bands are crossing the Fermi-level and thus can
participate in superconducting (SC) state.\cite{Haule08} Therefore
they are likely multi-band superconductors where the magnitude of
the energy gaps and even the sign of the order parameters can vary
between different bands.\cite{PRL-Kuroki, PRB-Bang, PRB-Dolgov}
Despite of the tremendous research efforts, there is an ongoing
debate about the basic superconducting properties, such as the
number of energy gaps and their magnitudes, which are
prerequisites for determining the order parameter symmetry and
ultimately the HTSC pairing mechanism.\cite{Node, PRB-Bang,
PRL-Kuroki, ARPES-el, ARPES-hole, Specific heat, muSR, NMR-1,
NMR-2, point contact, Wang-SC} Experimental evidence for at least
two different gaps has already been reported from angle-resolved
photoemission (ARPES), point contact spectroscopy, nuclear
magnetic resonance, and muon spin rotation
($\mu$SR).\cite{ARPES-el, ARPES-hole, Specific heat, muSR, NMR-1,
NMR-2, point contact} However, the reported gap values exhibit a
large variation from 2$\Delta/k_{B}T_{c} \approx$ 1.6 to 10 that
remains to be understood. Possible factors are: a variation
between the different compounds and as a function of doping and
structural changes, a strong dependence on the sample (surface)
quality, or the presence of even more than two gaps which the
various experimental techniques are probing with different
sensitivity. More detailed and accurate measurements of the
properties of the SC energy gap(s) are thus required.

Infrared (IR) spectroscopy is a powerful technique to investigate
the electronic gap structure of superconductors. While it does not
provide $k$-space resolved information such as ARPES, its large
probe depth ensures the bulk nature of the measured quantities and
its high energy resolution and powerful sum rules enable a
reliable determination of important physical parameters, such as
the gap magnitude and the plasma frequency of the SC condensate.
In the parent compound of the Ba 122
phase,(Ba,Sr)Fe$_{2}$As$_{2}$, the IR data revealed the spin
density wave gap(s) and related phonon
anomalies.\cite{Wang-undoped, phonon-Homes, Dressel} For the SC
single crystals the first report of the gap feature has been
published for hole doped Ba$_{1-x}$K$_{x}$Fe$_{2}$As$_{2}$ with
$T_{c}$ = 37 K.\cite{Wang-SC} In addition to a steep absorption
edge at 150 cm$^{-1}$ that is characteristic of a nearly isotropic
(nodeless) gap with 2$\Delta/k_{B}T_{c} \approx$ 6, they also
found evidence for a gap with 2$\Delta/k_{B}T_{c}\approx$ 8. A
similar upper value of 2$\Delta/k_{B}T_{c} \approx$ 8 has been
obtained from an ellipsometry study on polycrystalline
(Nd,Sm)FeAsO$_{1-x}$F$_{x}$.\cite{Dubroka} Recent studies on
electron doped BaFe$_{2-x}$Co$_{x}$As$_{2}$ single crystals
\cite{Heumen09, Wu09, Gorshunov10} also reported pronounced gap
edges around 30 - 50 cm $^{-1}$. These reports still vary
concerning the number and the magnitude of the SC energy gaps and
of the residual low-frequency conductivity.

In the following we present a detailed IR spectroscopy study of
electron doped BaFe$_{1.87}$Co$_{0.13}$As$_{2}$ single crystals
with $T_{c}$ = 24.5K. Our spectra reveal several SC-induced
features which yield further insight into the multigap nature of
SC. In particular, we show that a BCS-type model with three
isotropic energy gaps with 2$\Delta/k_{B}T_{c} \approx$ 3.1, 4.7,
and 9.2 describes the data rather well.

The BaFe$_{1.87}$Co$_{0.13}$As$_{2}$ single crystals were grown
from self-flux in glassy carbon crucibles and their chemical
composition was determined by energy dispersive x-ray spectroscopy
as described in Ref. 9. %\cite{Specific heat}.
A bulk SC transition of $T_{c} =$ 24.5 K was confirmed by
transport, dc magnetization, and $\mu$SR measurements.

The optical measurements were performed on freshly cleaved pieces
from a single growth batch. The temperature ($T$) dependent near
normal incidence reflectivity spectra, $R(\omega)$, for 35-5000
cm$^{-1}$ were measured with a Bruker 113v FT-IR spectrometer
utilizing the \textit{in situ} gold evaporation
technique.\cite{gold-evaporation} Additional ellipsometry
measurements were performed  at 350-8000 cm$^{-1}$ with a home
built rotating-analyzer ellipsometer attached to a Bruker 113v
(Ref. 23%\cite{ellipsometry-Bernhard}
) and at 6000 - 52000 cm$^{-1}$ with a commercial Woollam VASE
ellipsometer. All measurements were performed at least twice to
ensure their reproducibility. The ellipsometry data were converted
to normal incidence reflectivity data. The complex optical
constants were then calculated by Kramers-Kronig transformation
with proper extrapolations such that the directly measured
ellipsometric data were reproduced.

%%%%%%%%%%%%%%%%%%%%%%%%%%%%%%%%%%%%%%%%%%%%%%%

\begin{figure}
\includegraphics[width=8cm]{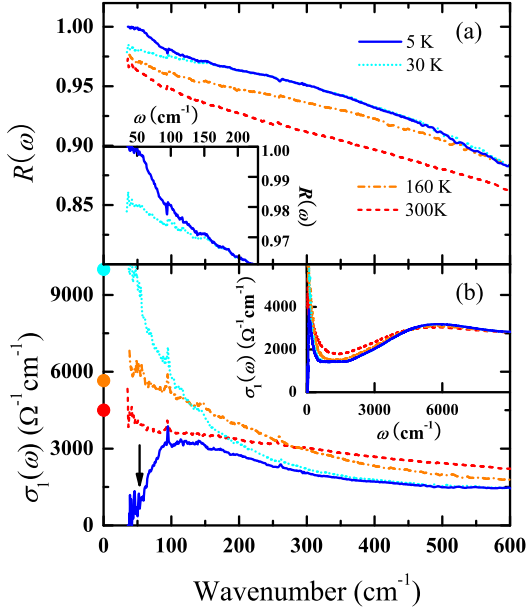}
\caption{Temperature dependence of the FIR reflectivity,
$R(\omega)$ (a) and optical conductivity $\sigma_1(\omega)$ (b).
Solid circles show $\sigma^{\rm{dc}}$ as reported for optimally
doped BaFe$_{1-x}$Co$_{x}$As$_{2}$ \cite{Alloul-transport}.
%The inset shows the conductivity spectra over a broader frequency range.
}
\label{fig.1}
\end{figure}

%%%%%%%%%%%%%%%%%%%%%%%%%%%%%%%%%%%%%%%%%%%%%%%%

Figure 1(a) shows the reflectivity, $R(\omega)$, in the
far-infrared (FIR) region at selected temperatures. The real part
of the optical conductivity, $\sigma_1(\omega)$, is
displayed in Fig. 1(b). %The inset shows a wider frequency range.
The overall shape and $T$- dependence of the spectra are similar
to previous reports.\cite{Wang-SC, Qazilbash, Heumen09, Wu09,
Gorshunov10, Perucchi10} In particular, the spectra contain a
pronounced Drude-peak at low frequency due to free carriers, which
narrows significantly as $T$ decreases in the normal state. The
extrapolated dc values, $\sigma^{\rm{dc}}$, compare well with
published transport data (solid circles).\cite{Alloul-transport}
The electronic response contains at least two more broad bands in
the mid-infrared (MIR) range that likely arise from interband
transitions.\cite{Wang-SC, Qazilbash} In addition, our data
contain two sharp features at 95 and 261 cm$^{-1}$ which
correspond to IR-active phonons. In the SC state at $T<T_{c}$, the
electronic response undergoes some characteristic changes below
about 300 cm$^{-1}$ which provide detailed information about the
SC energy gap(s) and the SC condensate.

%%%%%%%%%%%%%%%%%%%%%%%%%%%%%%%%%%%%%%%%%%%%%%%

\begin{figure}
\includegraphics[width=8cm]{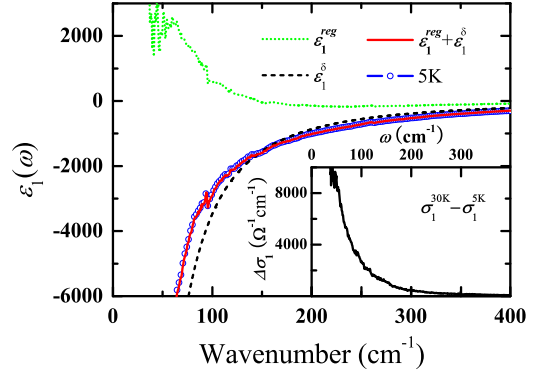}
\caption{Self-consist determination of the SC plasma frequency,
$\omega_{p,SC}$. The measured $\varepsilon_{1}^{\rm{5K}}$
(symbols) is shown to compare well with the expected spectrum
(solid line) that has been derived from the measured
$\varepsilon_{1}^{\rm{30K}}$ by subtracting the missing inductive
response due to
$\Delta\sigma_{1}$=$\sigma_{1}^{\rm{30K}}$-$\sigma_{1}^{\rm{5K}}$
(see the inset) to obtain $\varepsilon^{reg}$ (dotted line) and
adding the contribution $\varepsilon^{\delta}$ (dashed line) of
the SC delta function.} \label{fig.2}
\end{figure}

%%%%%%%%%%%%%%%%%%%%%%%%%%%%%%%%%%%%%%%%%%%%%%%%

First we discuss the determination of the plasma frequency of the
SC condensate, $\omega_{p,SC}$, which is detailed in Fig. 2. The
inset shows the SC-induced decrease in the regular part of the
conductivity,
$\Delta\sigma_{1}(\omega)$=$\sigma_{1}^{\rm{30K}}(\omega)$-$\sigma_{1}^{\rm{5K}}(\omega)$,
from which we obtain the missing spectral weight, $\Delta S =
\int_{0^{+}}^{\infty}$$\Delta\sigma_{1}(\omega)$$d\omega$. Its
magnitude falls off  steeply with increasing frequency and
essentially vanishes above 500 cm$^{-1}$ (see the inset of Fig.
2). According to the so-called Ferell-Glover-Tinkham sum rule,
$\Delta S$ is redistributed to a delta function at the origin
which accounts for response of the SC
condensate.\cite{Book-Tinkham} The analysis, using a low-frequency
extrapolation of the 30 K spectrum of $\sigma^{\rm{dc}}$ = 10000
$\Omega^{-1}$cm$^{-1}$ as reported in Ref. 26%\cite{Alloul-transport}
, yields $\Delta S \approx$ 9.1$\times 10^{5}$
$\Omega^{-1}$cm$^{-2}$ corresponding to $\omega_{p,SC} \approx$
5900 cm$^{-1}$ and a value of the magnetic penetration depth of
$\lambda_{ab}=\frac{c}{\omega_{p,SC}} \approx$ 270 nm. This agrees
reasonably well with $\lambda_{ab} \approx$ 230 nm as obtained
from $\mu$SR measurements on optimally doped
BaFe$_{1-x}$Co$_{x}$As$_{2}$.\cite{Bernhard-NJP, unpublished} The
self consistency of this analysis has been checked by inspecting
the corresponding changes in the real part of the dielectric
function, $\varepsilon_{1}(\omega)$. Figure 2 shows the comparison
between the measured $\varepsilon_{1}^{\rm{5K}}(\omega)$ (symbols)
and the corresponding spectrum (solid line) that is expected based
on the SC-induced spectral weight redistribution and, in
particular, the low frequency extrapolation that was used to
obtain $\omega_{p,SC}$. Specifically, from
$\varepsilon_{1}^{\rm{30K}}(\omega$) we subtracted the missing
contribution of the regular response due to
$\Delta\sigma_{1}(\omega)$=$\sigma_{1}^{\rm{30K}}(\omega)$-$\sigma_{1}^{\rm{5K}}(\omega)$
to obtain $\varepsilon^{reg}$ (dotted line) and added the
contribution $\varepsilon^{\delta}$ (dashed line) due to the SC
delta function. The excellent agreement between the symbols and
the solid line confirms that we performed a reasonable
low-frequency extrapolation of $\sigma_{1}$.

Next we show that valuable information about the SC energy gaps
can be obtained from the SC-induced changes in
$\sigma_{1}(\omega)$ and $R(\omega)$. First, we note that the
pronounced gap edge around 50  cm$^{-1}$ is the hallmark of an
(almost) isotropic energy gap whose magnitude does not far exceed
the normal state scattering rate, $\gamma$. In that case
$\sigma_{1}(\omega)$ exhibits a pronounced feature at 2$\Delta$
where it increases sharply from zero at $\omega<2\Delta$ to a
maximum around 4$\Delta$, before it eventually merges with the
normal state data at higher frequency.\cite{EPAPS, BCS-code}
Correspondingly, $R(\omega)$ remains unity at $\omega<2\Delta$ and
suddenly starts to decrease at $\omega>2\Delta$ yielding a
characteristic maximum in the reflectivity ratio
$R(T<T_c)/R(T\gtrsim T_c)$. \cite{EPAPS} In a multigap
superconductor, the corresponding features due to the larger gaps
are superimposed and thus more difficult to observe. Nevertheless,
as shown in the following, the gap magnitudes can still be
obtained based on an analysis with multiple independent isotropic
BCS-type superfluids.\cite{BCS-code}

%%%%%%%%%%%%%%%%%%%%%%%%%%%%%%%%%%%%%%%%%%%%%%%

\begin{figure}
\includegraphics[width=8cm]{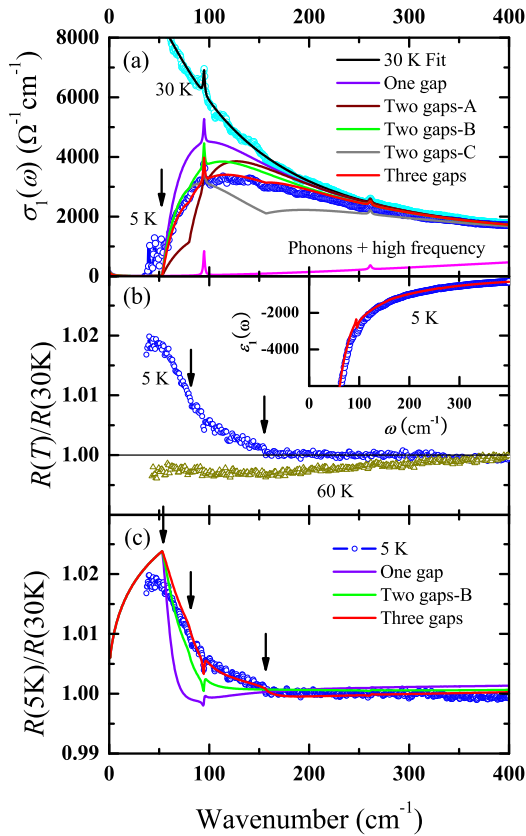}
\caption{Simulation with a BCS-type model of multiple isotropic
gaps. Comparison of the experimental spectra of the (a)
conductivity and (b)-(c) reflectivity ratio $R$($T$)/$R$(30K)
(symbols) with the simulated spectra (lines) as obtained by first
fitting the normal state response with a Drude-Lorentz-model and
then imposing one, two, or three independent isotropic BCS-type
energy gaps (parameters are given in Table I). Arrows mark clear
features due to the gap edges at 2$\Delta$. The inset shows the
corresponding experimental (symbol) and simulated (red line)
spectra of the dielectric function.} \label{fig.3}
\end{figure}

%%%%%%%%%%%%%%%%%%%%%%%%%%%%%%%%%%%%%%%%%%%%%%%%

At first we fitted the normal state spectra of
$\sigma_{1}^{\rm{30K}}(\omega$) and
$\varepsilon_{1}^{\rm{30K}}(\omega$) with a Drude-Lorentz model.
Since theory predicts that up to five bands are crossing the
Fermi-level, the low frequency response could be composed of as
many Drude bands. Nevertheless, we found that two Drude
oscillators (with parameters as given in Table I) are sufficient
for a reasonable description of the normal state data. Likely
these represent the electronlike and holelike bands, respectively.
Besides, we had to include two broad Lorentz oscillators centered
at about 1100 and 5400 cm$^{-1}$ to account for the electronic MIR
bands plus two narrow ones at 95 and 261 cm$^{-1}$ for the IR
active phonons. Their contribution is shown by the pink line in
Fig. 3(a) and assumed
not to change below $T_{c}$. Note that unlike in Ref. 19%\cite{Heumen09}
, our data yield no evidence for a sizeable electronic mode below
about 500 cm$^{-1}$ other than Drude terms. \cite{EPAPS}

First we demonstrate that the model of a single BCS-type gap fails
to describe our data.  In this case, the sharp edge in
$\sigma_{1}(\omega)$, as marked by the black arrow in Fig. 3(a)
defines the magnitude of the SC energy gap of $2\Delta \approx$ 53
cm$^{-1}$. Nevertheless, if we impose this gap value to the
Drude-bands as obtained at 30 K, the model provides a poor
description of the data since it largely overestimates the
magnitude of $\sigma_{1}(\omega>2\Delta)$, in particular, around
the maximum near 4$\Delta$.

Next we show that a good agreement with the experimental data can
be obtained with a three gap model. The signatures of the two
larger gaps are considerably weaker than the ones of the smallest
gap, but they can still be identified in the reflectivity ratio,
$R$(5K)/$R$(30K), in Fig. 3(b) which eliminates any additional
structures due to extrinsic features that may be introduced during
the thermal cycle in between the gold correction procedure. While
$R$(60K)/$R$(30K) shows no anomaly, $R$(5K)/$R$(30K) reveals two
noticeable kinks near 80 and 157 cm$^{-1}$ as marked by the black
arrows. These kinks are well separated from the narrow features
due to the phonon modes at 95 and 261 cm$^{-1}$. Therefore they
are supposed to be SC-induced ones, which should correspond to the
edges of the larger gaps.
%The edges of the larger gaps give rise to two noticeable kinks
%near 80 cm$^{-1}$ and 157 cm$^{-1}$ as marked by the black arrows.
%These kinks are well separated from the narrow features due to the
%phonon modes at 95 and 261 cm$^{-1}$ and they are clearly
%SC-induced since they are absent in $R$(60K)/$R$(30K).
Concerning the assignment of these gaps to the Drude bands (as
obtained from the 30 K spectrum), the best fits were obtained if
the gap structures at 53 and 157 cm$^{-1}$ (80 cm$^{-1}$) were
associated with the narrow (broad) Drude-band as detailed in Table
I. The quality of the fits was noticeably reduced if we changed
the assignment of the gaps at 53 and 80 cm$^{-1}$. Even for the
largest gap, that is associated with a rather small amount of
spectral weight, the comparison with the experiment become
significantly worse if we assigned it to the broad Drude band. The
best fit with this three gap model is shown by the red lines in
Fig. 3. It reproduces all the characteristic features of the
experimental data, including the long tail up to 150 cm$^{-1}$ in
$R$(5K)/$R$(30K). \cite{EPAPS, comment-ratio scaled}

Next we show that the best fits with a corresponding two-gap model
are significantly worse. Figure 3 (a) shows the results for three
representative assignments (A to C) of the gaps to the
Drude-bands. It highlights that the steep rise of
$\sigma_{1}(\omega)$ just above 53 cm$^{-1}$ can only be
reproduced if the smaller gap at 2$\Delta =$ 53 cm$^{-1}$ is
assigned to the narrow Drude-band as in models B and C.
Nevertheless, model B (C) severely underestimates (overestimates)
the magnitude of $\sigma_{1}(\omega)$ at higher frequency.
Furthermore, the two gap models fail to reproduce the pronounced
high frequency tail in $R$(5K)/$R$(30K) as is shown in Fig. 3(c)
for model B which yields the most reasonable fit with two gaps to
$\sigma_{1}(\omega)$. We note that alternative ways of fitting,
for example by describing the normal state spectrum with more than
two Drude-bands, did not improve these two-gap model fits. Also, a
strong coupling to a bosonic mode, while it might account for one
of the kink structures in $R$(5K)/$R$(30K), would further increase
$\sigma_{1}(\omega)$ above the boson energy and thus enhance the
discrepancy with the experimental data. \cite{strong-coupling}
Finally, we note that the inclusion of an additional low-frequency
electronic mode did not help to improve the quality of the fits
comparable to the one of the three gap model. \cite{EPAPS}

\begin{table}
\caption{Parameters for the simulations shown in Fig. 3.}
\label{Table.I}
\begin{ruledtabular}
\begin{tabular}{c|c|c|c|c}
  &\multicolumn{2}{c|}{Drude 1 ($\gamma$ = 90 cm$^{-1}$)}&\multicolumn{2}{c}{Drude 2 ($\gamma$ = 300
  cm$^{-1}$)}\\
  \cline{2-5}
  &2$\Delta$ (cm$^{-1}$)&$\omega_{p}^{2}$ (cm$^{-2}$)&2$\Delta$ (cm$^{-1}$)&$\omega_{p}^{2}$
  (cm$^{-2}$)\\
  \hline
  30 K&0&4.45$\times$10$^{7}$&0&4.3$\times$10$^{7}$\\
  \hline
  2 gaps-A&80&4.45$\times$10$^{7}$&53&4.3$\times$10$^{7}$\\
  \hline
  2 gaps-B&53&4.45$\times$10$^{7}$&80&4.3$\times$10$^{7}$\\
  \hline
  2 gaps-C&53&4.45$\times$10$^{7}$&157&4.3$\times$10$^{7}$\\
  \hline
  3 gaps&53&3.75$\times$10$^{7}$&80&4.3$\times$10$^{7}$\\
  \cline{2-5}
  &157&7.0$\times$10$^{6}$& &
\end{tabular}
\end{ruledtabular}
\end{table}

Our multigap analysis thus provides evidence for three different
energy gaps with 2$\Delta/k_{B}T_{c} \approx$ 3.1, 4.7, and 9.2 in
optimally doped BaFe$_{2-x}$Co$_{x}$As$_{2}$. These fall well into
the range of the reported values of 2$\Delta/k_{B}T_{c} \approx$
1.6 - 10 as obtained with various techniques. \cite{ARPES-el,
ARPES-hole, Specific heat, muSR, NMR-1, NMR-2, point contact} We
note that our optical data are compatible with the presence of a
fourth energy gap of magnitude 2$\Delta/k_{B}T_{c}<$ 3 or with gap
nodes. As long as this gap or the nodal states involves a
relatively small amount of spectral weight, we would not be able
to identify the corresponding signature in the spectrum of
$\sigma_{1}(\omega$) where the error bars become sizeable below 53
cm$^{-1}$. Some evidence (though not entirely conclusive) for an
energy gap with 2$\Delta<$40 cm$^{-1}$ or gap nodes is contained
in the spectrum of $R$(5)K/$R$(30). As shown in Fig. 3(c), it does
not exhibit the expected decrease below the lower gap edge at 53
cm$^{-1}$, instead it remains almost constant down to at least 40
cm$^{-1}$.

Concerning the assignment of these gaps to the different energy
bands our optical data provide only limited information. This
requires additional input concerning the momentum dependence for
example from ARPES. We also note recent theoretical work which
considers an order parameter with extended s-wave symmetry and
finds that impurity scattering can lead to a substantial
modification of the gap structures in the optical conductivity.
\cite{Carbotte10} Clearly, further experimental and theoretical
work is required to identify the number, magnitude and symmetry of
the energy gaps in these multi-band superconductors.

In summary, with combined IR reflection and ellipsometry
measurements we investigated the $T$-dependent optical in-plane
response of BaFe$_{1.87}$Co$_{0.13}$As$_{2}$ single crystals with
$T_{c}$ = 24.5 K.We showed that the spectra display clear
SC-induced changes that are characteristic of at least three
different energy gaps. In particular, we showed that they can be
well accounted for in terms of a BCS-type multigap model with
three isotropic gaps with 2$\Delta/k_{B}T_{c} \approx$ 3.1, 4.7,
and 9.2. In addition, we determined the SC plasma frequency and
the corresponding low-$T$ value of the in-plane magnetic
penetration depth of 270 nm.

\begin{acknowledgments}
Part of the work was performed at the IR beamline of the ANKA
synchrotron at FZ Karlsruhe, D, where we appreciate the technical
support of  Y.L. Mathis. We acknowldge financial support by the
Schweizer Nationalfonds (SNF) under Grant No. 200020-119784 and
No. 200020-129484, and the NCCR project MaNEP, and by the Deutsche
Forschungsgemeinschaft (DFG) under Grant No. BE2684/1-3 in FOR538.

\end{acknowledgments}

%\newpage

%\newpage


\begin{references}

\bibitem{Kamihara08} Y. Kamihara et al., J. Am. Chem. Soc. {\bf 130}, 3296 (2008).

\bibitem{Haule08} K. Haule, J. H. Shim, and G. Kotliar, Phys. Rev. Lett. {\bf 100}, 226402 (2008).

\bibitem{PRB-Dolgov} O.V. Dolgov, I.I. Mazin, D. Parker, and
A.A. Golubov, Phys. Rev. B {\bf 79}, 060502(R) (2009).

\bibitem{PRL-Kuroki} K. Kuroki et al., Phys. Rev. Lett. {\bf 101}, 087004 (2008).

\bibitem{PRB-Bang} Yunkyu Bang and Han-Yong Choi, Phys. Rev. B {\bf 78}, 134523 (2008).

\bibitem{Node} R. T. Gordon et al., Phys. Rev. Lett. {\bf 102},
127004 (2009); J. K. Dong et al., Phys. Rev. Lett. {\bf 104},
087005 (2010).

\bibitem{ARPES-el}  K. Terashima et al., Proceedings of the National Academy of
Sciences of the USA (PNAS) 106, 7330 (2009).

\bibitem{ARPES-hole} H. Ding et al., Euro. Phys. Lett. {\bf 83}, 47001 (2008).

\bibitem{Specific heat} F. Hardy et al., Phys. Rev. B {\bf 81}, 060501(R) (2010).

\bibitem{muSR} T.J. Williams et al., Phys. Rev. B {\bf 80}, 094501
(2009).

\bibitem{point contact} P. Szab\'{o} et al., Phys. Rev. B {\bf 79}, 012503
(2009).

\bibitem{NMR-1} M. Yashima et al., J. Phys. Soc. Jpn. {\bf 78} 103702 (2009).

\bibitem{NMR-2} K. Matano, G.L. Sun, D.L. Sun, C.T. Lin, and Guo-qing Zheng,
EPL {\bf 87} 27012 (2009).

\bibitem{Wang-SC} G. Li et al., Phys. Rev. Lett. {\bf 101}, 107004 (2008).

\bibitem{phonon-Homes} A. Akrap et al., Phys. Rev. B {\bf 80}, 180502(R) (2009).

\bibitem{Wang-undoped} W.Z. Hu et al., Phys. Rev. Lett. {\bf 101}, 257005 (2008).

\bibitem{Dressel} D. Wu et al., Phys. Rev. B {\bf 79}, 155103 (2009).

\bibitem{Dubroka} A. Dubroka et al., Phys. Rev. Lett. {\bf 101}, 097011  (2008).

\bibitem{Heumen09} E. van Heumen et al., arXiv:0912.0636.

\bibitem{Wu09} D. Wu et al., arXiv:0912.0849.

\bibitem{Gorshunov10} B. Gorshunov et al., Phys. Rev. B {\bf 81}, 060509(R) (2010).

\bibitem{gold-evaporation} C. C. Homes, M. Reedyk, D. A. Cradles, and T. Timusk, Appl. Opt. {\bf 32}, 2976 (1993).

\bibitem{ellipsometry-Bernhard} C. Bernhard, J. Huml\'{i}cek, and B. Keimer, Thin Solid Films {\bf 455-456}, 143 (2004).

\bibitem{Perucchi10} A. Perucchi et al., arXiv: 1003.0565.

\bibitem{Qazilbash} M.M. Qazilbash et al., Nature Physics {\bf 5}, 647 (2009).

\bibitem{Alloul-transport} F. Rullier-Albenque, D. Colson, A. Forget, and H.
Alloul, Phys. Rev. Lett. {\bf 103}, 057001 (2009).

\bibitem{Book-Tinkham} Michael Tinkham, {\it Introduction to
Superconductivity} (McGraw-Hill, Singapore, 1996).

\bibitem{Bernhard-NJP} C. Bernhard et al., New J. Phys. {\bf 11}, 055050 (2009).

\bibitem{unpublished} Unpublished $\mu$SR data on the same crystals.

%\bibitem{Basov Science and PRB on Tl-2201} D.N. Basov et al., Science {\bf 283}, 49 (1999);
%A.S. Katz et al., Phys. Rev. B {\bf 61}, 5930 (2000).

\bibitem{EPAPS} See supplementary material at http://link.aps.org/
supplemental/10.1103/PhysRevB.81.214508 for specific details of
the data analysis.

\bibitem{BCS-code} W. Zimmermann, E.H. Brandt, M. Bauer, E. Seider, and L. Genzel,
Physica C {\bf 183}, 99 (1991).

\bibitem{comment-ratio scaled} The calculated reflectivity ratio
is scaled up by a factor of 1.002, which gives a good agreement at
high frequency where the spectrum is featureless. This discrepancy
might originate from the numerical code, which generates slightly
larger spectral weight than the model Drude term, or from
imperfection of the fit to the normal state (Ref. 30).%\cite{EPAPS}

\bibitem{strong-coupling} Sang Boo Nam, Phys. Rev. {\bf 156}, 487
(1967).

\bibitem{Carbotte10} J.P. Carbotte and E.Schachinger, Phys. Rev. {\bf B 81}, 104510
(2010).

\end{references}
\end{document}